\documentclass[12pt,preprint]{aastex}

\usepackage{epsfig}
 
\begin{document} 
\title{Discovery of 14 radio pulsars in a survey of the Magellanic Clouds}
\author{R. N. Manchester\altaffilmark{1}, G. Fan\altaffilmark{2,3},
  A. G. Lyne\altaffilmark{4}, V. M. Kaspi\altaffilmark{3} \and 
F. Crawford\altaffilmark{5}}
\altaffiltext{1}{Australia Telescope National Facility, CSIRO, PO Box 76,
        Epping NSW 1710, Australia}
\altaffiltext{2}{Physics Department, University of Hong Kong, Pokfulham Rd, Hong Kong, 
China}
\altaffiltext{3}{Department of Physics, Rutherford Physics Building, McGill University, 
Montreal, QC H3A 2T8, Canada}
\altaffiltext{4}{University of Manchester, Jodrell Bank Observatory, Macclesfield, 
Cheshire SK11 9DL, UK}
\altaffiltext{5}{Department of Physics, Haverford College, Haverford, PA 19041-1392}
\email{dick.manchester@csiro.au}

\begin{abstract} 
A systematic survey of the Large and Small Magellanic Clouds for radio
pulsars using the Parkes radio telescope and the 20-cm multibeam
receiver has resulted in the discovery of 14 pulsars and the
redetection of five of the eight previously known spin-powered pulsars
believed to lie in the Magellanic Clouds. Of the 14 new discoveries,
12 are believed to lie within Clouds, three in the Small Cloud and
nine in the Large Cloud, bringing the total number of known
spin-powered pulsars in the Clouds to 20. Averaged over all positions
within the survey area, the survey had a limiting flux density of
about 0.12 mJy. Observed dispersion measures suggest that the mean
free electron density in the Magellanic Clouds is similar to that in
the disk of our Galaxy. The observed radio luminosities have little or
no dependence on pulsar period or characteristic age and the
differential luminosity function is consistent with a power-law slope
of $-1$ as is observed for Galactic pulsars.
\end{abstract}


\section{Introduction}
The Large and Small Magellanic Clouds (LMC and SMC) are our
nearest-neighbour galaxies and, to date, the only galaxies other than
our own which have detectable pulsars. The first systematic survey for
radio pulsars in the Magellanic Clouds was by \citet{mhah83} who used
a receiver with center frequency 645 MHz and bandwidth 25 MHz on the
Parkes 64-m radio telescope to search 7 square degrees of the
LMC. This survey uncovered the first known extra-galactic pulsar, PSR
B0529$-$66, which has a period of 0.975 s and a dispersion measure
(DM) of 103 cm$^{-3}$ pc which clearly places it outside the Galactic
free-electron layer. This survey was subsequently extended by
\citet{mmh+91} to cover both the LMC and the SMC with higher
sensitivity using a receiver with center frequency 610 MHz, a
bandwidth 60 MHz in each of two orthogonal polarizations and 5-ms
sampling. With an observation time of 5000 s per pointing, the
limiting mean flux density at 610 MHz was 0.5 to 0.8 mJy for pulsars
with periods greater than about 500 ms and DMs of 100 cm$^{-3}$ pc or
less. Four pulsars were discovered, one in the SMC, two in the LMC and
one foreground pulsar. The SMC pulsar, now known as PSR J0045$-$7319,
was shown by \citet{kjb+94} to be in a 51-day orbit round a massive
star, optically identified as a 16th-magnitude B star. The
identification was subsequently confirmed by the detection of orbital
Doppler shifts in spectral lines from the B-star, showing that the
star has a mass of 8 -- 10 M$_\sun$ \citep{bbs+95}.

Many young pulsars emit pulsed emission in the X-ray band and
sometimes detection is easier in this band than at radio
wavelengths. Two such pulsars which are clearly located in the LMC are
known. The first, PSR B0540$-$69, was discovered by \citet{shh84}
using data from the {\it Einstein X-ray Observatory}. 
With a pulse period of 50.2 ms and a characteristic age of just 1670
years, it has properties very similar to those of the Crab pulsar and is
located at the center of the supernova remnant (SNR) 0540$-$693 in the
LMC \citep{msk93}. This pulsar was detected in the radio band by
\citet{mml+93} at 640 MHz with a very broad pulse and mean flux
density of about 0.4 mJy. Giant radio pulses at 1400 MHz were detected
from PSR B0540$-$69 by \citet{jr03} and subsequently the 1400-MHz mean
pulse emission was observed with a flux density of just 24 $\mu$Jy by
\citet{jrmz04}. Another young Magellanic Cloud pulsar, PSR
J0537$-$6910, was detected by \citet{mgz+98} using observations with
the {\it Rossi X-ray Timing Explorer}. This pulsar lies within the SNR
N~157B in the LMC and has the shortest period, 16.1 ms, of any known
young (unrecycled) pulsar. Its characteristic age is 5000 years, close
to the estimated age of N~157B. Radio searches for pulsed emission
from this pulsar have been unsuccessful, with the best limit on the
mean flux density, 10 $\mu$Jy at 1400 MHz, being set by
\citet{cmj+05}.

A decade after the \citet{mmh+91} search, \citet{ckm+01} exploited the
high survey efficiency of the 13-beam Parkes 20-cm multibeam receiver
\citep{swb+96,mlc+01} to undertake a higher-sensitivity survey of the
SMC. A total area of $\sim 6.7$ square degrees was covered in 12
pointings of the instrument, with 0.25-ms sampling interval and
8400 s observation time per pointing giving a 1400-MHz limiting flux
density of about 0.08 mJy for pulsars with periods greater than 50 ms
and DMs less than 200 cm$^{-3}$ pc. For comparison, the 610-MHz
limiting flux density of the \citet{mmh+91} survey corresponds to a
1400-MHz limiting flux density of about 0.12 to 0.18 mJy assuming a
spectral index of $-1.8$, typical for most pulsars
\citep{mkkw00a}. These limiting flux densities apply to a detection at
the beam center and do not take into account variations of telescope
gain over the surveyed region.  The \citet{ckm+01} survey found two
previously unknown pulsars, one of which is believed to be associated
with the SMC. In a separate search for pulsed emission from PSR
J0537$-$6910, a pulsar, PSR J0535$-$6935, was serendipitously
discovered.

These discoveries brought the total number of known pulsars associated
with the Magellanic Clouds to eight, seven of which are radio
emitting, two of which are in the SMC and six of which were discovered
in radio surveys. In this paper we report on an extensive radio survey
of both the LMC and SMC for pulsars. Like the \citet{ckm+01} survey of
the SMC, this survey used the multibeam receiver on the Parkes 64-m
radio telescope. It has a comparable sensitivity to the \citet{ckm+01}
survey but covers a wider area and is sensitive to higher DMs. The
survey was very successful with 14 pulsars discovered, 12 of which are
believed to be in the Magellanic Clouds. In \S2 we describe the survey
and timing observations, \S3 gives details of the new discoveries and
the implications of the results are discussed in \S4.

\section{Observations and analysis procedures}
\begin{figure}[ht]
\plottwo{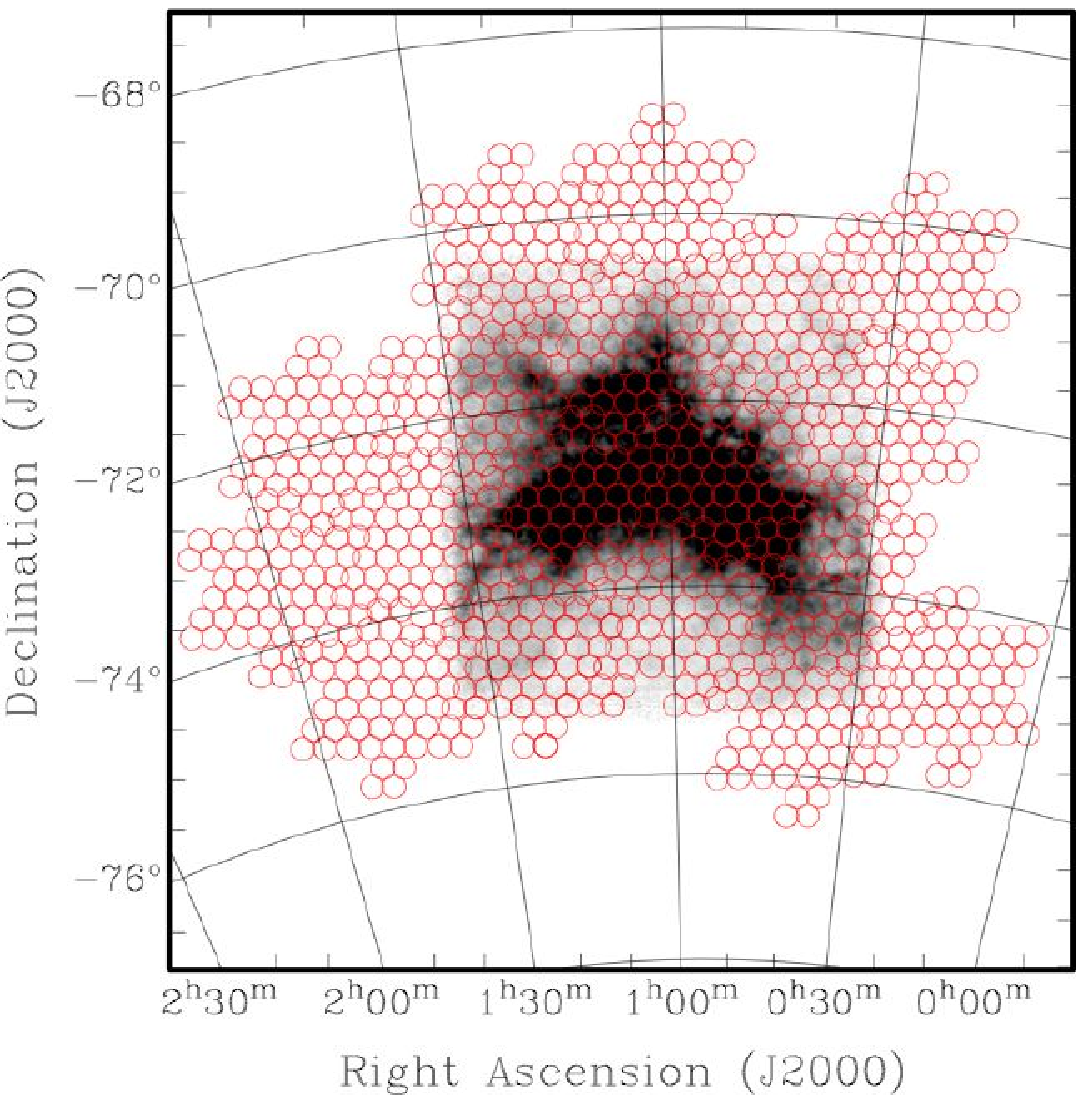}{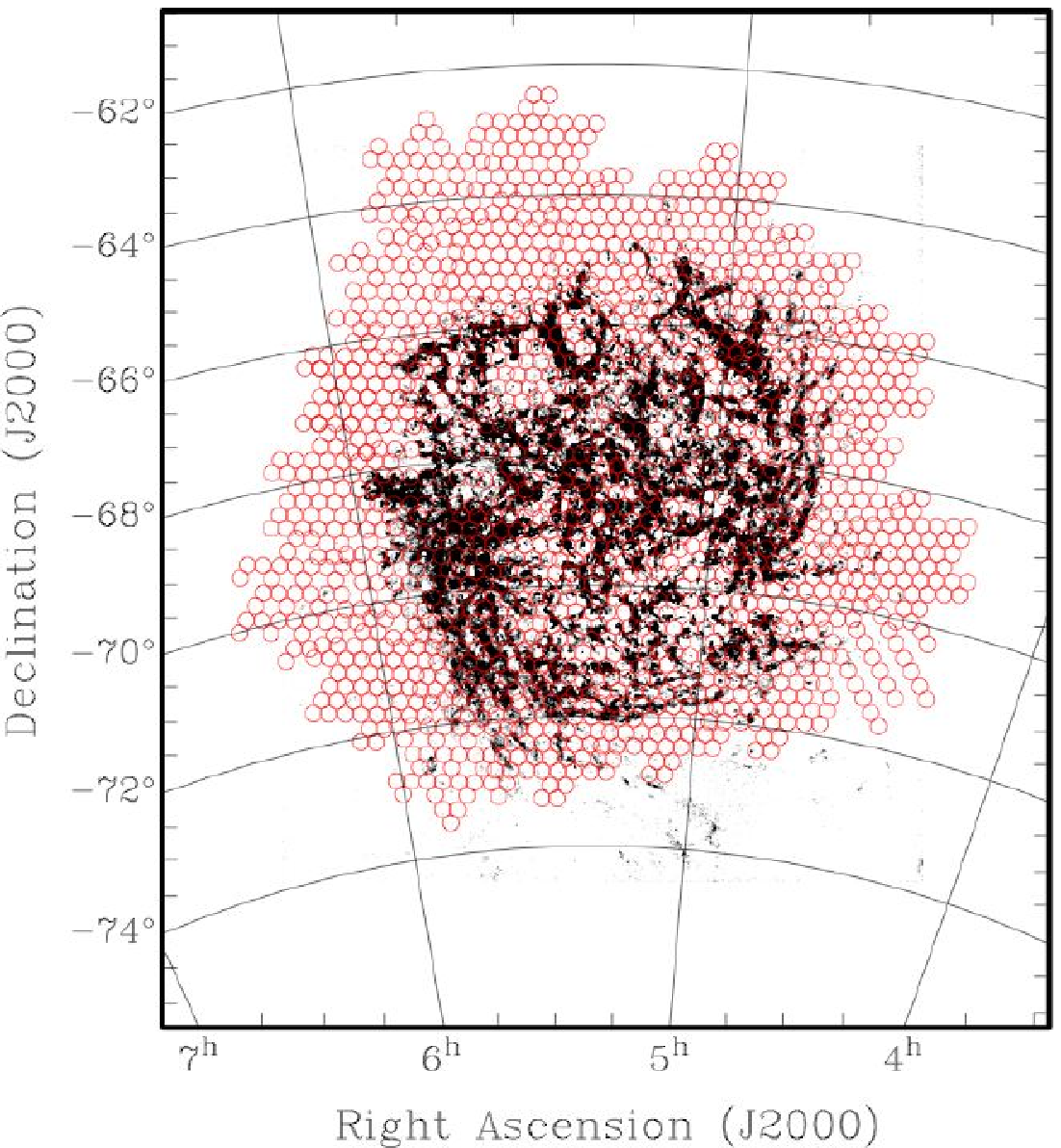}
\caption{Distribution of survey beams across the SMC (left) and LMC
  (right) overlaid on a greyscale of the neutral hydrogen distribution
  from \citet{ssd+99} and \citet{ksd+03} for the SMC and LMC
  respectively. }\label{fg:beams}
\end{figure}

The survey observations were made using the 13-beam multibeam receiver
\citep{swb+96} on the Parkes 64-m radio telescope in several sessions
between 2000 May and 2001 November. The 13 beams lie
in a double-hexagon pattern around a central beam which allows
complete sky coverage (with adjacent beams overlapping at the
half-power points) with interleaved pointings. A brief description of
the observing system follows; for more details see \citet{mlc+01}. Each
beam has cryogenic receivers for two orthogonal polarizations centered at 1374
MHz. For each polarization, the received 288-MHz band is split into 96
3-MHz channels. Detected signals from each channel are summed in
polarization pairs, high-pass filtered, integrated for 1 ms, one-bit
sampled and written to Digital Linear Tape for subsequent
analysis. Each pointing was observed for 8400 s, giving $2^{23}$
samples per observation. A total of 73 pointings (949 beams) was
observed for the SMC and 136 pointings (1768 beams) for the LMC. The
distribution of these beams overlaid on images of the neutral hydrogen
in the SMC and LMC is shown in Figure~\ref{fg:beams}. 

In most respects, off-line processing followed the procedures outlined
by \citet{ckm+01} and \citet{mlc+01} and was carried out on Sun
workstations at the ATNF and at McGill University. Compared to the
\citet{ckm+01} survey of the SMC, a larger DM range was searched in
the present survey. A total of 102 DMs up to 277 cm$^{-3}$ pc were
searched with 1-ms time resolution, a further 42 DMs were then
searched for each of 2~ms, 4~ms and 8~ms time resolution, with maximum
DMs of approximately 550 cm$^{-3}$ pc, 1100 cm$^{-3}$ pc and 2200
cm$^{-3}$ pc, respectively. For most periods and DMs, the sensitivity
of this survey was comparable to that of \citet{ckm+01}, with a
limiting flux density of about 0.08 mJy for an assumed 5\% duty
cycle. Because of the four-times-longer sampling interval, the
sensitivity for short-period pulsars ($P\la50$ ms) is reduced compared
to the earlier survey. These limiting flux densities apply to a pulsar
at the center of the central beam of the multibeam receiver. In
general, pulsars will lie some distance from the center of the nearest
beam. Furthermore, outer beams have a lower gain than the central
beam. Following \citet{mlc+01}, we estimate the limiting mean flux
density of the survey averaged over all positions within the tiled
region to be approximately 0.12 mJy.

The data analysis resulted in a large number of candidate pulsars
which were graded according to signal-to-noise ratio (S/N) and absence
of interference. The better candidates were reobserved using a grid
pattern of observations with the center beam of the multibeam system
at the nominal position and $9\arcmin$ north, south, east and west,
typically for 3000~s per pointing. The data from these observations
were searched over ranges of period and DM about the nominal
values. Detection of a pulsar signal in one or more pointings gave
confirmation of the candidate and detection in two or more pointings
also allowed determination of an improved position for the pulsar.

Following confirmation of a pulsar, a program of timing observations
commenced in order to improve the position, period and DM, and to
determine the period derivative. These timing observations used the
center beam of the multibeam receiver and the same recording system as
was used for the survey. The resulting data were folded at the
topocentric period to form mean pulse profiles which were further
analysed using the {\sc psrchive} \citep{hvm04} pulsar data analysis
package to produce pulse times of arrival (TOAs) for each observation
(where there was a detectable pulse). These were analysed using {\sc
tempo}\footnote{See http://www.atnf.csiro.au/research/pulsar/tempo}
with the DE200 Solar-System ephemeris \citep{sta90} and a weighted
fit. Typically, the timing observations spanned 400 -- 600 days with
more than 20 observations per pulsar. In most cases, DMs were
determined by splitting the observed 20-cm band into four subbands and
measuring TOAs separately for each subband. In a few cases the pulsar
was also detected using a 70-cm receiver which has a bandwidth of 32
MHz centered at 436 MHz; combination of the 20-cm and 70-cm TOAs
resulted in an improved DM for these pulsars. For each pulsar, several
observations having high S/N were folded at two and three times the
nominal period in order to uncover detection by the survey analysis
software of a harmonic of the true pulse frequency --- this procedure
revealed two pulsars that had mis-identified periods from the survey
analysis.

\section{Results}
The procedures described above resulted in the detection of 14 previously
unknown pulsars, three toward the SMC and 11 toward the
LMC. Table~\ref{tb:par} lists the J2000 right ascension and
declination obtained from the timing analysis, the DM and its
``z-component'', DM $\sin|b|$, where $b$ is the Galactic latitude,
which is useful for determining if the pulsar is within or outside the
Galactic electron layer, the mean flux density (averaged over the
pulse period) at 1400 MHz and the pulse width at 50\% of the peak
amplitude. Table~\ref{tb:ppar} gives other parameters derived from the
timing analysis, namely the barycentric pulse period, $P$, its first time
derivative, $\dot P$, the epoch of the period in Modified Julian Days,
the time-span of the data used in the timing fit, the number of TOAs fitted
and the post-fit rms timing residual. Estimated uncertainties given in
parentheses after parameter values are twice the {\sc tempo} rms
errors and refer to the last quoted digit. Mean flux densities were
estimated from the area under grand average profiles formed by summing
data from all timing observations (whether or not the pulsar was
detectable) using the derived timing model to phase-align the
profiles.

PSR J0451$-$67 was detected just twice in two years of
timing observations and a coherent timing solution was not
possible. Grid observations were also unsuccessful, so that the quoted
position is that of the discovery beam and the uncertainties are the
half-power beam radius. The period given in Table~\ref{tb:ppar} was
obtained by splitting the best single observation into five parts of
equal duration and obtaining TOAs for each part; the quoted error
includes a contribution from the position uncertainty. A limit on the
period derivative was obtained by differencing the derived period and
that from the discovery observation on MJD 51816. The mean flux
density was estimated by taking the value for the best detection (on
MJD 52817) and dividing by $N^{1/2}$ where $N$ is the total number of
timing observations ($\sim 20$).

\begin{deluxetable}{llllcrr}
\tablecaption{Parameters for Pulsars Discovered in this Survey\label{tb:par}}
\tablehead{
\colhead{PSR} & \colhead{RA (J2000)} & \colhead{Dec. (J2000)} & \colhead{DM} & 
\colhead{DM$\sin|b|$} & \colhead{S$_{1400}$} & \colhead{W$_{50}$} \\
 & \colhead{(h m s)}  & \colhead{($\degr$ $\arcmin$ $\arcsec$)} &
 \colhead{(cm$^{-3}$ pc)} & \colhead{(cm$^{-3}$ pc)} & \colhead{(mJy)} & \colhead{(ms)}
}
\startdata
J0045$-$7042  & 00:45:25.69(17) & $-$70:42:07.1(13)  & 70(3)   & 50.7 &   0.11 & 19  \\
J0111$-$7131  & 01:11:28.77(9)  & $-$71:31:46.8(6)   & 76(3)   & 54.2 &   0.06 & 13  \\
J0131$-$7310  & 01:31:28.51(3)  & $-$73:10:09.34(13) & 205.2(7)&141.6 &  0.15 & 4.8 \\
J0449$-$7031  & 04:49:05.67(5)  & $-$70:31:31.7(3)   & 65.83(7)\tablenotemark{*}& 38.2 &   0.14 & 7.9 \\
J0451$-$67    & 04:51:50(70)    & $-$67:18(7)        & 45(1)   & 26.6 &   $\lesssim 0.05$ & 5.5 \\
J0456$-$7031  & 04:56:02.5(3)   & $-$70:31:06.6(12)  & 100.3(3)\tablenotemark{*}& 57.5 &   0.05 & 8  \\
J0457$-$6337  & 04:57:07.79(8)  & $-$63:37:30.4(9)   & 27.5(10)& 16.4 &   0.18 & 36 \\
J0511$-$6508  & 05:11:56.50(2)  & $-$65:08:36.5(3)   & 25.66(8)\tablenotemark{*}& 14.6 &   0.70 & 12  \\
J0519$-$6932  & 05:19:46.917(12)& $-$69:32:23.48(7)  & 119.4(5)& 65.5 &   0.32 & 4.1  \\
J0522$-$6847  & 05:22:23.06(8)  & $-$68:47:02.2(3)   &126.45(7)\tablenotemark{*}& 69.2 &   0.19 & 12  \\
J0532$-$6639  & 05:32:59.51(6)  & $-$66:39:37.3(5)   & 69.3(18)& 37.2 &   0.08 & 9  \\
J0534$-$6703  & 05:34:36.17(10) & $-$67:03:48.8(8)   & 94.7(12)& 50.6 &	 0.08 & 25  \\
J0543$-$6851  & 05:43:52.71(11) & $-$68:51:25.3(9)   & 131(4)  & 67.9 &   0.22 & 58  \\
J0555$-$7056  & 05:55:01.85(12) & $-$70:56:45.6(6)   & 73.4(16)& 36.8 &   0.21 & 27  \\
\enddata
\tablenotetext{*}{70-cm and 20-cm data used for DM determination.}
\end{deluxetable}						  

\begin{deluxetable}{lllllrc}
\tablecaption{Period Parameters for Pulsars Discovered in this Survey\label{tb:ppar}}
\tablehead{
\colhead{PSR} & \colhead{$P$} & \colhead{$\dot P$} & \colhead{Epoch} & 
\colhead{Data Span} & \colhead{$N_{\rm TOA}$} & \colhead{Residual}\\
 & \colhead{(s)}  & \colhead{($10^{-15}$)} & \colhead{MJD} & 
 \colhead{MJD} &  & \colhead{(ms)}
}
\startdata
J0045$-$7042 & 0.63233580002(6)  & 2.49(3)   & 52407.0 & 52189 -- 52642 & 17 & 1.54  \\
J0111$-$7131 & 0.68854151164(5)  & 7.092(4)  & 52369.0 & 52041 -- 52695 & 15 & 1.14  \\
J0131$-$7310 & 0.348124045581(7) & 1.7614(7) & 52335.0 & 52038 -- 52631 & 24 & 0.40  \\
J0449$-$7031 & 0.479163971291(14)& 3.371(4)  & 52465.0 & 52233 -- 52695 & 20 & 0.51  \\
J0451$-$67\tablenotemark{a} & 0.24545429(7) & $<2$ & 52817.0 & 52817
-- 52818 & 5 & 0.80  \\
J0456$-$7031 & 0.80013207321(10) & 36.70(3)  & 52560.0 & 52316 -- 52804 & 15 & 2.25  \\
J0457$-$6337 & 2.49701169613(19) & 0.21(3)  & 52465.0 & 52189 -- 52739 & 8 & 0.75  \\
J0511$-$6508 & 0.322061827468(8) & 0.206(4) & 52409.0 & 52189 -- 52628 & 27 & 0.29  \\
J0519$-$6932 & 0.263211634568(3) & 0.6968(5) & 52466.0 & 52237 -- 52695 & 26 & 0.20  \\
J0522$-$6847 & 0.67453190906(3)  & 17.727(14)& 52503.0& 52311 -- 52695 & 16 & 0.29  \\
J0532$-$6639 & 0.64274275093(3)  & 5.354(13) & 52462.0 & 52259 -- 52664 & 14 & 0.58  \\
J0534$-$6703 & 1.81756503106(17) & 425.05(6) & 52410.0 & 52195 -- 52624 & 16 & 0.65  \\
J0543$-$6851 & 0.70895418575(6)  &  3.94(3) & 52410.0 & 52195 -- 52624 & 16 & 1.32  \\
J0555$-$7056 & 0.82783808575(7)  &  5.96(3) & 52408.0 & 52190 -- 52624 & 16 & 0.84  \\
\enddata
\tablenotetext{a}{See text.}
\end{deluxetable}

\begin{figure}[ht]
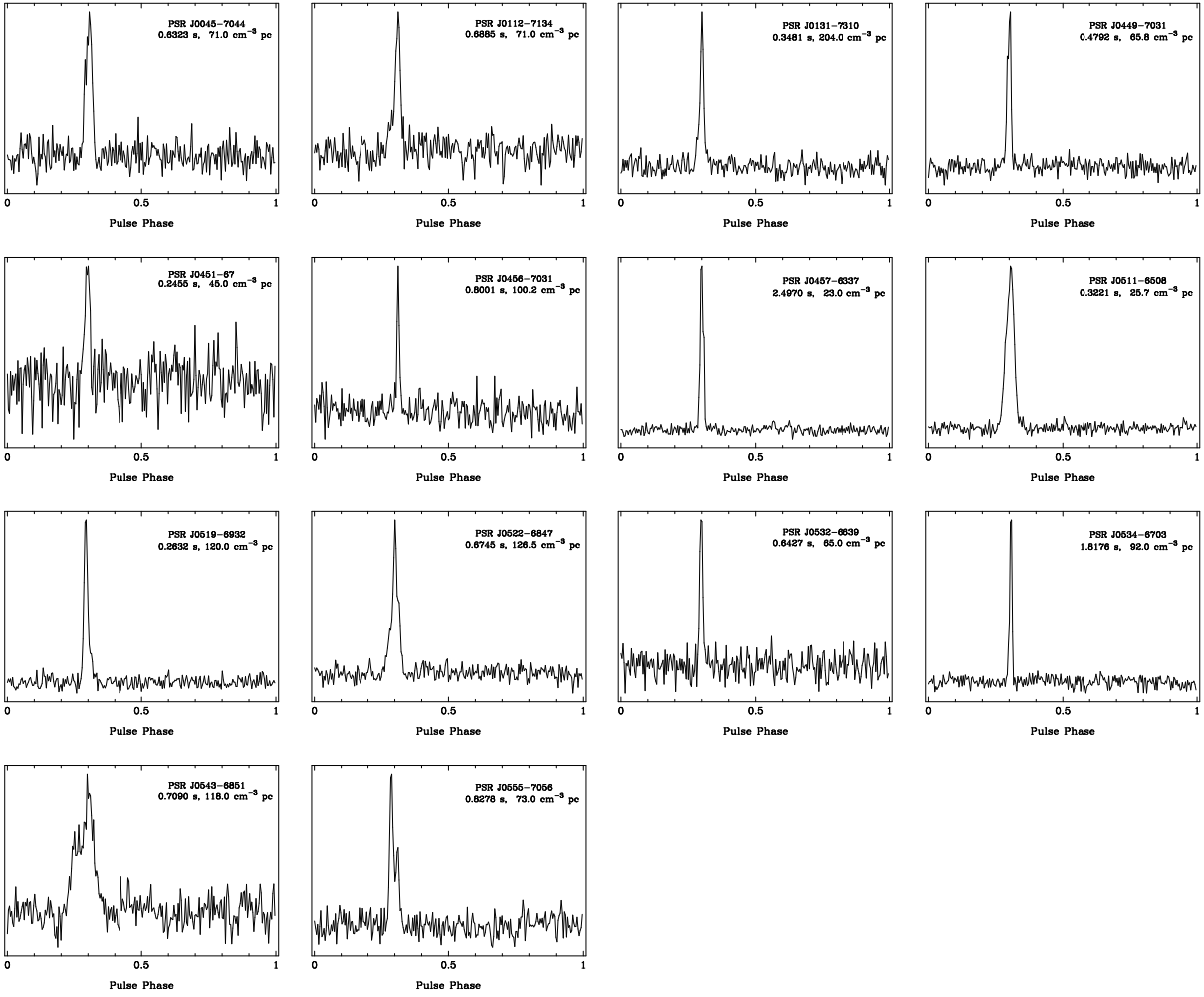

\begin{tabular}{cccc}
\mbox{\psfig{file=f2a.eps,width=30mm,angle=270}} &
\mbox{\psfig{file=f2b.eps,width=30mm,angle=270}} &
\mbox{\psfig{file=f2c.eps,width=30mm,angle=270}} &
\mbox{\psfig{file=f2d.eps,width=30mm,angle=270}} \\
\mbox{\psfig{file=f2e.eps,width=30mm,angle=270}} &
\mbox{\psfig{file=f2f.eps,width=30mm,angle=270}} &
\mbox{\psfig{file=f2g.eps,width=30mm,angle=270}} &
\mbox{\psfig{file=f2h.eps,width=30mm,angle=270}} \\
\mbox{\psfig{file=f2i.eps,width=30mm,angle=270}} &
\mbox{\psfig{file=f2j.eps,width=30mm,angle=270}} &
\mbox{\psfig{file=f2k.eps,width=30mm,angle=270}} &
\mbox{\psfig{file=f2l.eps,width=30mm,angle=270}} \\
\mbox{\psfig{file=f2m.eps,width=30mm,angle=270}} &
\mbox{\psfig{file=f2n.eps,width=30mm,angle=270}} 
\end{tabular}
\caption{Mean total intensity pulse profiles at 1400 MHz for the 14
  pulsars discovered in this survey. The pulsar period and dispersion
  measure are given on each plot below the name.}\label{fg:prf}
\end{figure}

Mean pulse profiles at 1400 MHz for the 14 pulsars are shown in
Figure~\ref{fg:prf}; the whole pulse period is shown in each case.
These profiles were obtained by summing data from all observations of
a given pulsar where there was a detectable pulse and then correcting
for the effects of the high-pass filter in the digitization system;
see \citet{mlc+01} for details. Profiles were rotated to place the
pulse peak at phase 0.3 for display purposes. It is notable that, in
most cases, the mean pulse profiles are narrow and relatively simple
in form. The median duty cycle ($W_{50}/P$) for these pulsars is just
0.015 (compared to 0.029 for the Galactic population) and the largest
(for PSR J0543$-$6851) is only 0.082. Only two pulsars, PSRs
J0543$-$6851 and J0555$-$7056, have clear evidence for multiple
components in the mean profile.

Five of the seven previously known radio pulsars in the Magellanic
Clouds were detected in the survey: PSRs J0045$-$7319, J0113$-$7220,
J0455$-$6915, J0502$-$6617 and J0529$-$6652. PSR J0540$-$6919 was well
below our detection threshold and PSR J0535$-$6935 has an estimated
1400-MHz flux density of 0.05 mJy; neither pulsar was detected in the
survey. However, in an attempt to obtain a timing solution for PSR
J0535$-$6935, about 50 observations with the beam centered on the
nominal position and mostly with a duration of about 3 hours were made
between late 1998 and 2004 resulting in about 25 detections. The
period and DM giving the highest S/N for the pulse were determined for
each observation by searching over a small range in each
parameter. The derived periods showed a clear trend over the six
years. Although a coherent timing solution could not be found, a fit
of a straight line to the seven points having S/N greater than 8.0
gave $P$ = 0.20051133(2) s at MJD 52200 and $\dot P = 11.5(3)\times
10^{-15}$ with an assumed position of R.A. 05$^{\rm h}$~35$^{\rm
m}$~00$^{\rm s}$, Dec. $-$69\degr~35\arcmin~00\arcsec. These values
are consistent with the period parameters quoted by \citet{ckm+01} for
this pulsar. Averaging the DM values for the seven high-S/N detections
gave a mean value of $93.7\pm0.4$ cm$^{-3}$ pc.

\section{Discussion}
Of the 14 pulsars listed in Table~\ref{tb:par}, two have values of
DM~$\sin |b|$ significantly less than 25 cm$^{-3}$ pc. As discussed by
\citet{ckm+01} the distribution of DM~$\sin |b|$ for Galactic pulsars
extends to about this value, suggesting that these pulsars, PSRs
J0457$-$6337 and J0511$-$6508, probably lie within the disk of our
Galaxy. A third pulsar, PSR J0451$-$67, has a DM~$\sin |b|$ value
marginally above this limit, so its association is not clear. However,
only nine of the roughly 1500 known Galactic-disk pulsars have
DM~$\sin |b|$ greater than the value for PSR J0451$-$67.\footnote{Data
obtained from the ATNF Pulsar Catalogue, Version 1.24
(\url{http://www.atnf.csiro.au/research/pulsar/psrcat}; Manchester et
al. 2005).}\nocite{mhth05} Furthermore, of the eight globular
clusters containing known pulsars and with Galactic $z$-distances
greater than 3 kpc, only two have pulsars with DM~$\sin |b|$ values
greater than that for PSR J0451$-$67. We therefore conclude that it is
most probable that PSR J0451$-$67 lies in the LMC, most likely on the
near side. All other pulsars in Table~\ref{tb:par} have significantly
higher values of DM~$\sin |b|$ and are almost certainly associated
with the Magellanic Clouds.

\begin{figure}[ht]
\plotone{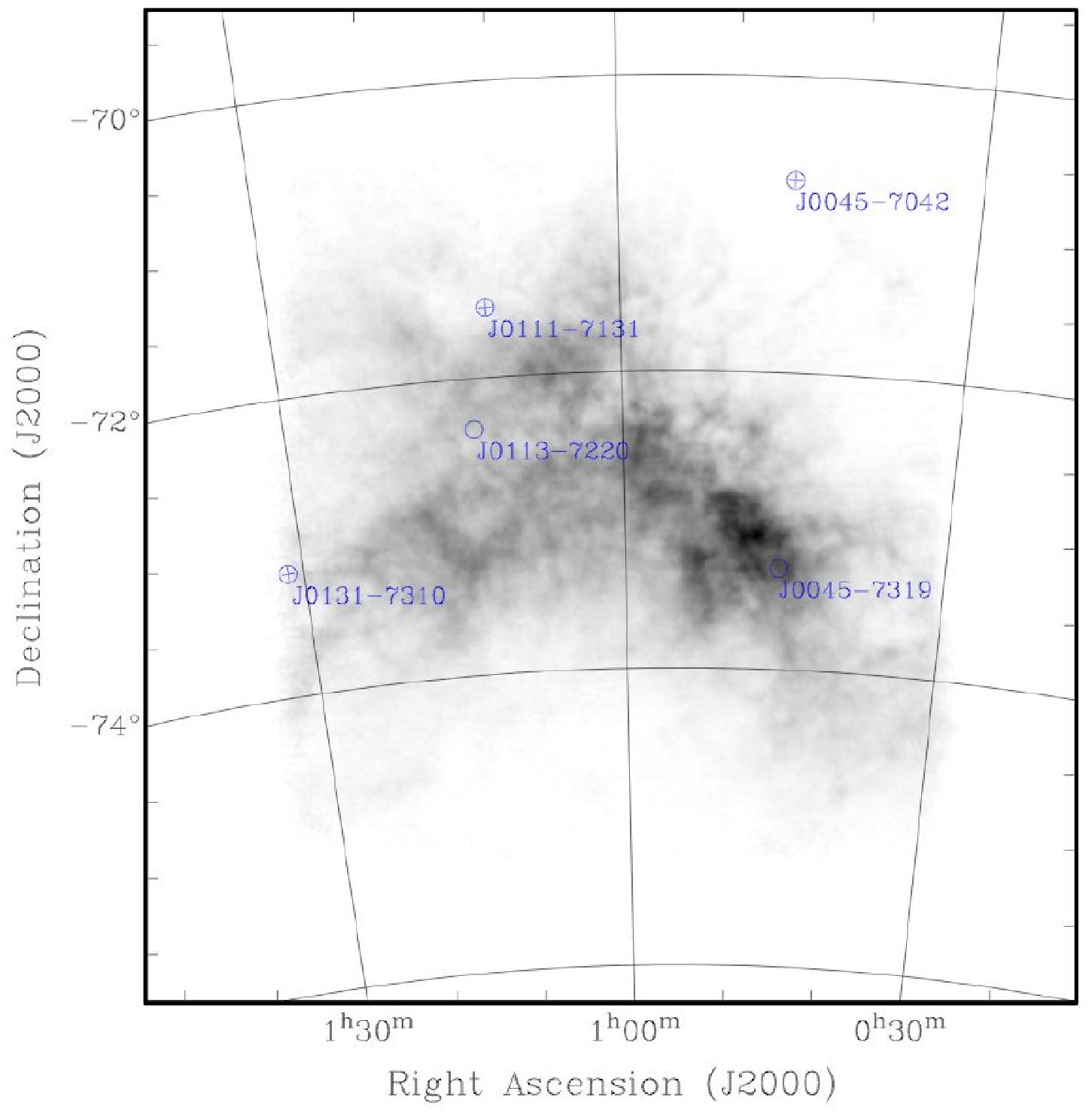}
\caption{Location of pulsars discovered in the SMC survey, marked by a
  cross, overlaid on an image of the neutral hydrogen in the SMC
  \citep{ssd+99}. Pulsars believed to be associated with the LMC,
  including previously known pulsars, are marked by a
  circle.}\label{fg:smc_psrs}
\end{figure}

Figure~\ref{fg:smc_psrs} shows the positions of the three SMC pulsars
from Table~\ref{tb:par} together with the two previously known SMC
pulsars plotted on a greyscale image of the HI in the SMC. Two of the
newly discovered pulsars are within the HI distribution (at least in
projection) but the other, PSR J0045$-$7042 is outside the main HI
region. PSR J0131$-$7310 lies on the eastern edge of the imaged region
but is notable for having the highest DM and DM~$\sin |b|$ of any of
the Magellanic Cloud pulsars by a considerable
margin. Figure~\ref{fg:beams} shows that a considerable area further
to the east of PSR J0131$-$7310 was searched, but no pulsars were
detected there.

\begin{figure}[ht]
\plotone{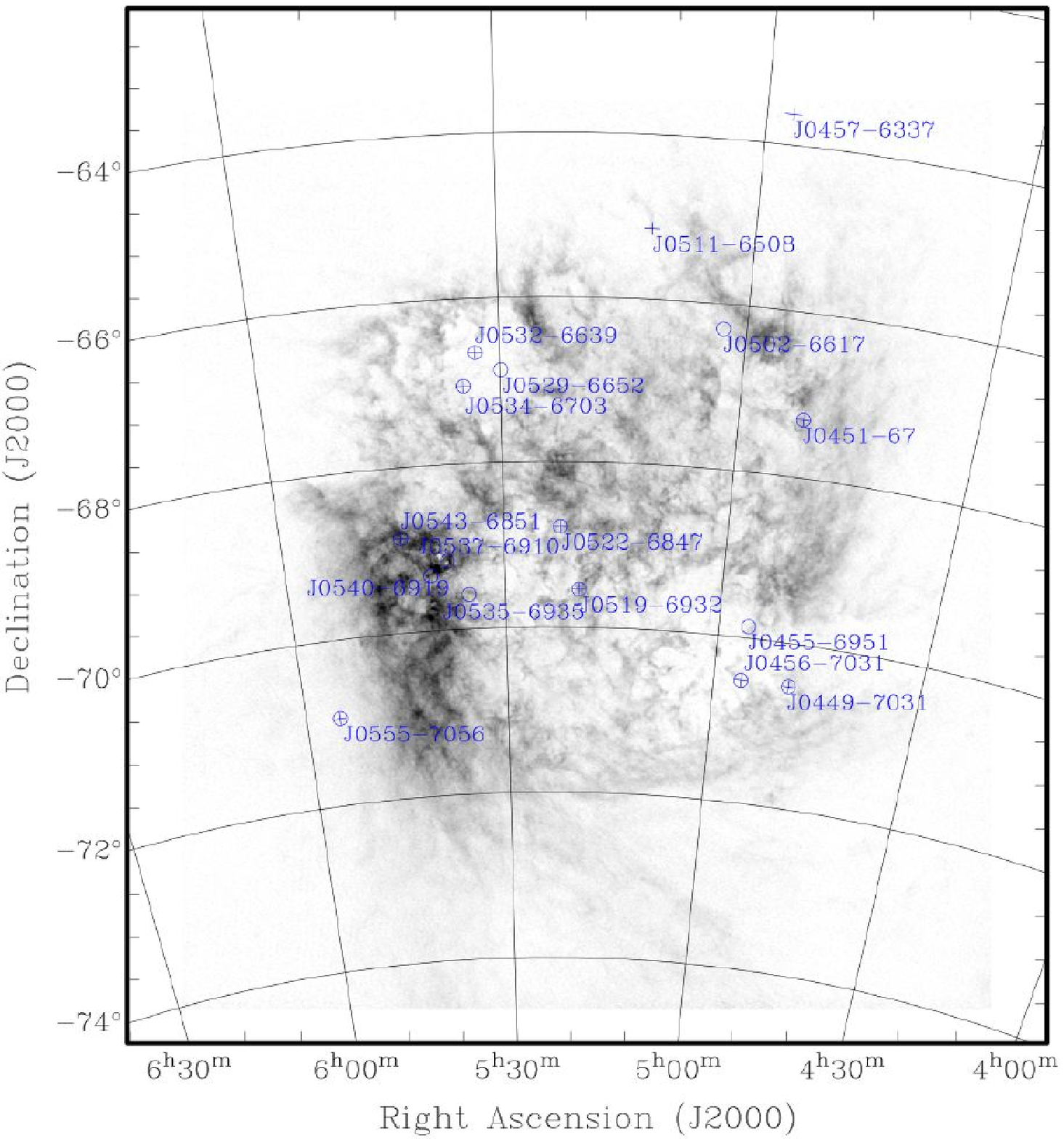}
\caption{Location of pulsars discovered in the LMC survey, marked by a
  cross, overlaid on an image of the neutral hydrogen in the LMC
  \citep{ksd+03}. Pulsars believed to be associated with the LMC,
  including previously known pulsars, are marked by a
  circle.}\label{fg:lmc_psrs}
\end{figure}

Positions of newly discovered pulsars in the direction of the LMC and
previously known pulsars believed to be associated with the LMC are
shown in Figure~\ref{fg:lmc_psrs}. All the pulsars believed to be associated with
the LMC lie within the HI distribution. The two foreground pulsars lie
outside the main HI distribution to the north.

No millisecond pulsars were discovered in the survey. This is not
surprising given the 1-ms sampling interval employed and the generally
lower radio luminosity of millisecond pulsars. However, two very
long-period pulsars were discovered (Table~\ref{tb:ppar}). One of
these, PSR J0457$-$6337, has a small period derivative and hence a
very large age, $\sim 190$ Myr. Although we do not believe this pulsar
to be associated with the LMC, it has a significant DM~$\sin |b|$
(Table~\ref{tb:par}) and must be located in the Galactic halo. Both of
the commonly used models for the Galactic electron density
\citep{tc93,cl02} place this (and PSR J0511$-$6508) at
or beyond the outer edge of the Galactic electron disk with
$z$-distances in excess of 1 kpc. The other long-period pulsar
discovered in the survey, PSR J0534$-$6703, is clearly associated with
the LMC and is notable for its very large period derivative
(Table~\ref{tb:ppar}). This pulsar is relatively young ($\tau_c =
P/2\dot P \sim 68$ kyr) and has a very strong implied surface dipole
magnetic field, $B_s = 3.2\times 10^{19} (P\dot P)^{1/2} \;{\rm G} \sim
2.8\times 10^{13}$ G. Only ten radio pulsars have a greater surface
dipole magnetic field based on their observed period and period
derivative.

Table~\ref{tb:prop} lists the 20 known spin-powered pulsars believed
to be associated with the Magellanic Clouds, 12 of which were
discovered in the present survey. The table lists the pulse period
$P$, the DM (except for PSR J0537$-$6910 which has not been detected
at radio wavelengths), the $\log_{10}$ of the characteristic age
$\tau_c$ in years, the mean pulsed flux density at 1400 MHz
$S_{1400}$, the monochromatic radio luminosity at 1400 MHz $L_{1400} =
S_{1400} d^2$, where $d$ is the pulsar distance, the discovery
reference and the reference for the flux density measurement. The
distance $d$ is assumed to be 50 kpc in all cases; this is reasonable,
since the actual distances are not reliably known. In any case,
uncertainties in the luminosities are contributed to, if not dominated
by, uncertainties in the mean flux densities.

\begin{deluxetable}{llccccccc}
\tablecaption{Properties of Magellanic Cloud Pulsars\label{tb:prop}}
\tablehead{
\colhead{JName} & \colhead{BName} & \colhead{$P$} & \colhead{DM} & \colhead{log($\tau_c$)} &
\colhead{$S_{1400}$} & \colhead{$L_{1400}$} & \colhead{Disc.} & \colhead{$S_{1400}$} \\
 & & \colhead{(s)} & \colhead{(cm$^{-3}$ pc)} & & \colhead{(mJy)} & \colhead{(mJy kpc$^2$)} &
\colhead{Ref.} & \colhead{Ref.} }
\startdata
J0045$-$7042 &  \nodata  & 0.6323 &  70.0 & 6.60 & 0.11   & 275 & 1 & 1  \\
J0045$-$7319 &  \nodata  & 0.9262 & 105.4 & 6.51 & 0.3  	& 750 & 2 & 3  \\
J0111$-$7131 &  \nodata  & 0.6885 &  76.0 & 6.18 & 0.06  	& 150 & 1 & 1  \\
J0113$-$7220 &  \nodata  & 0.3258 & 125.4 & 6.02 & 0.4  	& 1000& 3 & 3  \\
J0131$-$7310 &  \nodata  & 0.3481 & 205.2 & 6.49 & 0.15  	& 375 & 1 & 1  \\
J0449$-$7031 &  \nodata  & 0.4791 &  65.8 & 6.35 & 0.14  	& 350 & 1 & 1  \\
J0451$-$67   &  \nodata  & 0.2454 &  45   & $>6.2$ & $\lesssim 0.05$& $\lesssim 125$ & 1 & 1  \\
J0455$-$6951 &  B0456$-$69 & 0.3204 &  94.8 & 5.69 & 0.25\tablenotemark{*} & 625 & 2 & 2  \\
J0456$-$7031 &  \nodata  & 0.8001 & 100.3 & 5.53 & 0.05  	& 125 & 1 & 1  \\
J0502$-$6617 &  B0502$-$66 & 0.6912 &  68.9 & 5.67 & 0.25\tablenotemark{*} & 625 & 2 & 2  \\
J0519$-$6932 &  \nodata  & 0.2632 & 119.4 & 6.77 & 0.32  	& 800 & 1 & 1  \\
J0522$-$6847 &  \nodata  & 0.6745 & 126.4 & 5.78 & 0.19  	& 475 & 1 & 1  \\
J0529$-$6652 &  B0529$-$66 & 0.9757 & 103.2 & 6.00 & 0.3  	& 750 & 4 & 3  \\
J0532$-$6639 &  \nodata  & 0.6427 &  69.3 & 6.28 & 0.08  	& 200 & 1 & 1  \\
J0534$-$6703 &  \nodata  & 1.8175 &  94.7 & 4.83 & 0.08  	& 200 & 1 & 1  \\
J0535$-$6935 &  \nodata  & 0.2005 &  89.4 & 5.44 & 0.05  	& 125 & 3 & 3  \\
J0537$-$6910 &  \nodata  & 0.0161 &\nodata& 3.69 &$<0.01$	& $<25$ & 5 & 6  \\
J0540$-$6919 &  B0540$-$69 & 0.0505 & 146.5 & 3.22 & 0.024 	& 60  & 7 & 8  \\
J0543$-$6851 &  \nodata  & 0.7089 & 131.0 & 6.45 & 0.22  	& 550 & 1 & 1  \\
J0555$-$7056 &  \nodata  & 0.8278 &  73.4 & 6.34 & 0.21  	& 525 & 1 & 1  \\
\enddata
\tablenotetext{*}{From $S_{610}$ assuming a spectral index of $-1.8$}
\tablecomments{References: (1) This paper; (2) \citet{mmh+91}; (3)
  \citet{ckm+01}; (4) \citet{mhah83}; (5) \citet{mgz+98}; (6)
  \citet{cmj+05}; (7) \citet{shh84}; (8) \citet{jrmz04}.}
\end{deluxetable}

Based on the observed pulsar positions and DMs, we can estimate the
free-electron content of the two Clouds. For the SMC and LMC
separately, we first subtract a Galactic contribution, $25/\sin |b|$
cm$^{-3}$ pc from the DM for each pulsar and compute the mean value of
the Cloud DMs, obtaining $\langle DM\rangle = 80.9$ cm$^{-3}$ pc for
the SMC and 49.5 cm$^{-3}$ pc for the LMC. If we assume pulsars are
uniformly distributed through a roughly spherical volume for each
Cloud, the total integrated electron content across the Cloud is twice
the mean value or approximately 160 cm$^{-3}$ pc ($\sim 5\times
10^{20}$ cm$^{-2}$) for the SMC and 100 cm$^{-3}$ pc ($\sim 3\times
10^{20}$ cm$^{-2}$) for the LMC. We also estimate the mean electron
density in each cloud by computing the rms dispersion of the Cloud
DMs, 48.2 cm$^{-3}$ pc and 26.8 cm$^{-3}$ pc for the SMC and LMC
respectively, and dividing this by the (one-dimensional) rms
dispersion of the pulsar spatial coordinates in the right ascension
and declination directions (assuming a mean distance of 50 kpc and
that the offsets in the two directions are independent), 1025 pc and
1480 pc respectively. This procedure gave mean electron densities of
0.047 cm$^{-3}$ for the SMC and 0.018 cm$^{-3}$ for the LMC. We are
somewhat limited by small-number statistics in the conclusions that we
can draw from these results. For example, the relatively large values
for the SMC may simply be a consequence of PSR J0131$-$7310 lying
behind a dense HII region. Never-the-less, the results indicate that
the mean free electron density in the Magellanic Clouds is similar to
that in the disk of our Galaxy, possibly with a somewhat higher value
in the SMC and a somewhat lower value in the LMC. These results show
that the Magellanic Clouds have a much higher proportion of stars
contributing to ionization of the interstellar medium compared to the
Galaxy, consistent with the much higher per-unit-mass star-formation
rates in the Clouds \citep[e.g.,][]{ggs03}.

\begin{figure}[ht]
\plottwo{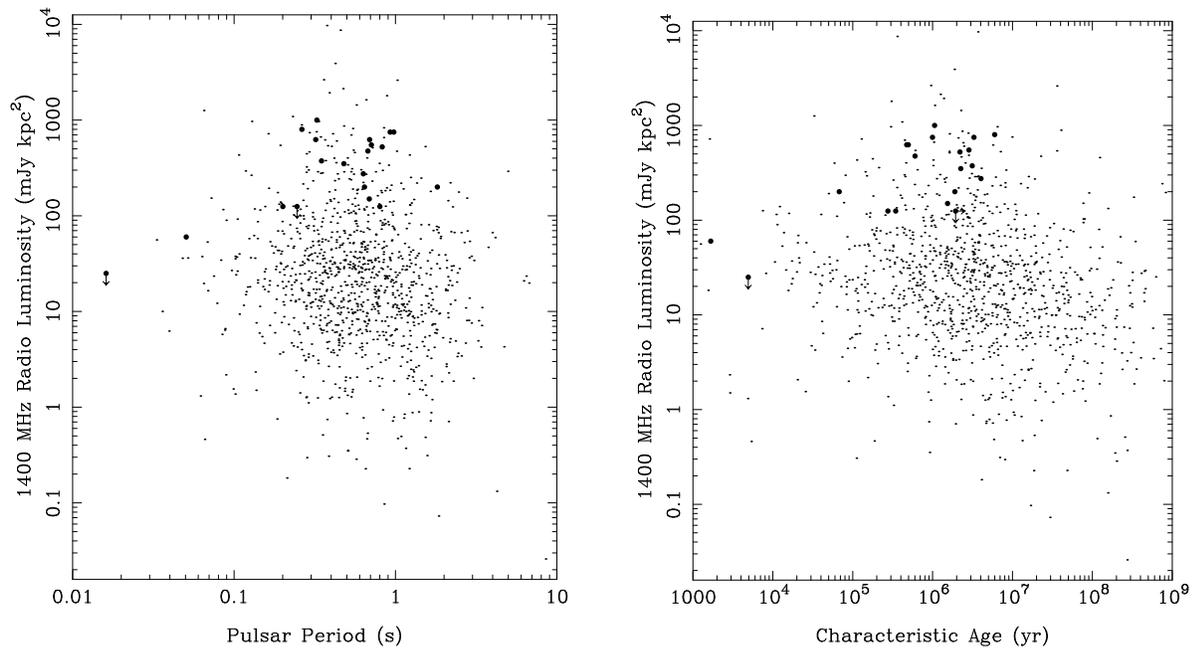}{f5b.eps}
\caption{Plot of 1400 MHz radio luminosity for non-recycled pulsars in
the Galactic disk (dots) and Magellanic Clouds (filled circles) versus
pulsar period (left) and characteristic age (right). Parameters for
Galactic pulsars were obtained from the ATNF Pulsar
Catalogue.}\label{fg:lum_distr}
\end{figure}

Despite the lack of an electron density model for the Magellanic
Clouds, the relative distance uncertainties are small for the
Magellanic Cloud pulsars compared to those in our Galaxy. Therefore
Magellanic Cloud pulsars give a more reliable estimation of the
luminosity and its dependence on various parameters than can be
obtained from Galactic pulsars. Figure~\ref{fg:lum_distr} shows the
1400 MHz radio luminosity for pulsars in the Galactic disk and the
Magellanic Cloud plotted against pulsar period (left) and
characteristic age (right). The approximately 1000 Galactic pulsars
plotted in these figures exclude apparently recycled pulsars,
identified by having $B_s < 3\times 10^{10}$~G, as well as pulsars in
globular clusters.  The Magellanic Cloud pulsars are typically of
higher radio luminosity than Galactic pulsars. This is simply a result
of the limited sensitivity of radio surveys combined with the greater
distance of the Magellanic Cloud pulsars.

Although the number of Magellanic Cloud pulsars remains relatively
small, they do cover significant ranges of both pulse period and
age. The maximum observed period is 1.82~s and more than half the ages
are greater than $10^6$ years. Figure~\ref{fg:lum_distr} shows that
there is little or no significant dependence of radio luminosity on
either pulsar period or age. This remains true even if PSR
J0537$-$6910 and PSR J0540$-$6919, neither of which was discovered in
radio surveys, are disregarded. Indeed, if anything, there is a weak
positive correlation of radio luminosity with characteristic age!
This suggests that the emitted radio power is not greatly dependent on
pulsar period or age and that other factors dominate the observed
luminosity. It is worth noting that, assuming radiation into a beam of
angular width equal to the observed pulse width, the radio luminosity
is a tiny fraction, typically $10^{-4}$, of the spin-down luminosity.

The true radio luminosity may be substantially underestimated if only
the outer part of the radio beam sweeps across the
Earth. Unfortunately, there is little known about the pulse
polarization of the Magellanic Cloud pulsars, so no estimates of
offsets of our viewing angle from the beam center are
available. However, experience with Galactic pulsars shows that, even
when these data are available, our poor understanding of radio beam
shapes precludes making reliable corrections for these offsets
\citep[cf.][]{tm98}. The narrow and relatively simple pulse profiles
observed for most of the pulsars suggests that the observed emission
is core-dominated, but this remains to be confirmed by polarization
and spectral measurements. As mentioned in \S3, the median duty cycle
($w$) for pulsars detected in this survey, 0.015, is considerably less
than the median value for the Galactic population, 0.029 (0.028 if
millisecond pulsars are excluded). Since the Magellanic Cloud pulsars
are all of high radio luminosity, this might suggest a correlation
between luminosity and duty cycle, but no such correlation exists in
the known Galactic population. It is more likely that the low median
duty cycle is a selection effect resulting from the fact that the
limiting flux density is proportional to $w^{1/2}$ for small $w$. It
is easier to detect pulsars of a given mean flux density if they have
a small duty cycle. This effect is significant when a large fraction
of the sample is close to the detection threshold.

The two shortest-period and youngest known Magellanic Cloud pulsars
were both first detected at X-ray wavelengths. Both have low radio
luminosities, with only an upper limit for PSR J0537$-$6910. It is
quite possible that the radio beam for both of these pulsars misses or
largely misses the Earth, implying that the X-ray beam either covers a
wider range of pulsar latitude than the radio beam or is less patchy.

\begin{figure}[ht]
\plotone{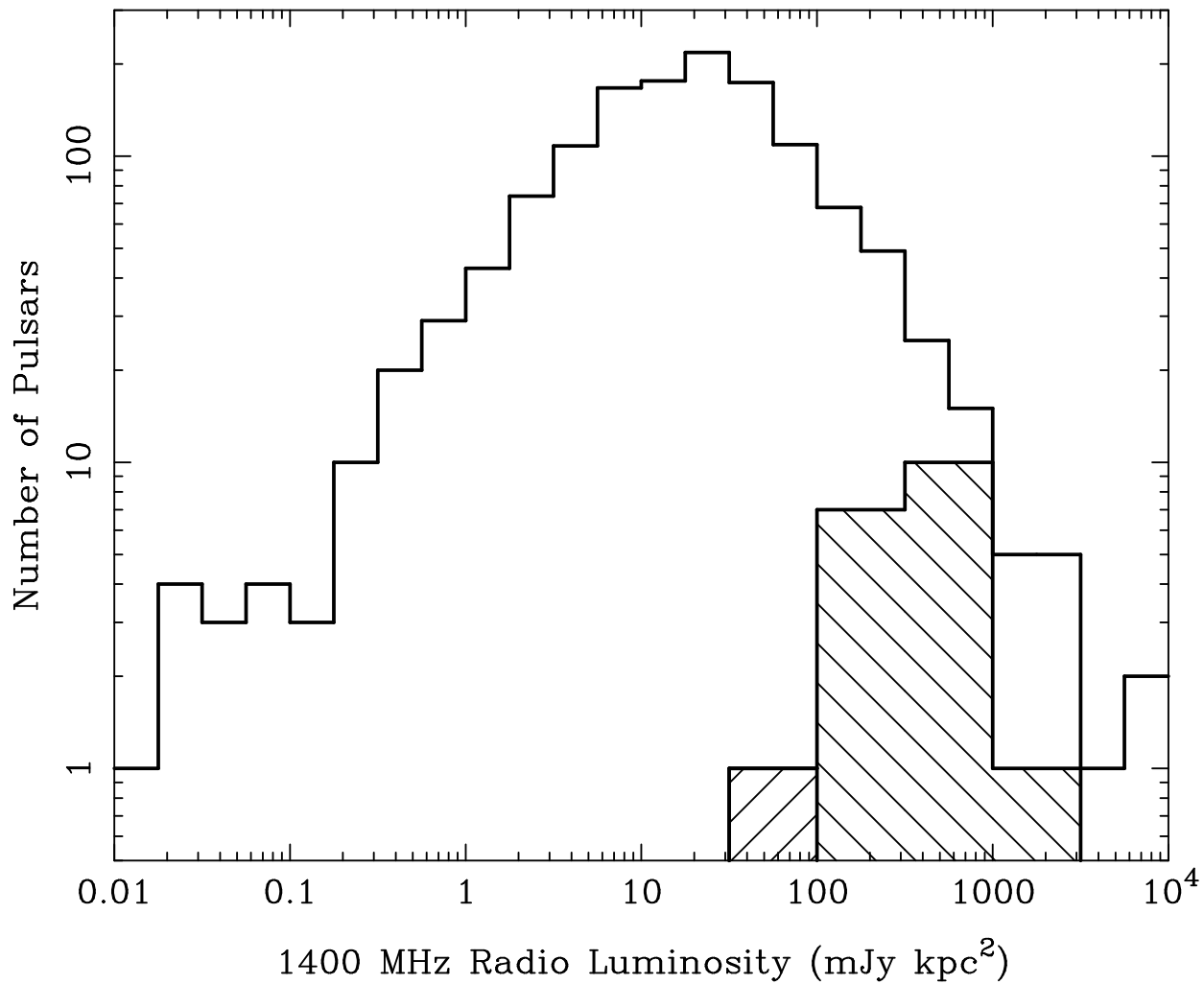}
\caption{Observed distribution of 1400 MHz radio luminosities for the
  Magellanic Cloud pulsars (hatched) and Galactic disk pulsars. The
  Magellanic Cloud pulsar with the lowest luminosity is PSR
  J0540$-$6919 (shown with different hatching) which was discovered at
  X-ray wavelengths and subsequently detected at radio
  wavelengths.}\label{fg:lumfn}
\end{figure}

Figure~\ref{fg:lumfn} shows the distribution of observed radio
luminosities for Galactic and Magellanic Cloud pulsars. This figure
clearly illustrates the effect of selection against low flux densities
in radio pulsar surveys. For luminosities above about 30 mJy kpc$^2$
for Galactic pulsars and 300 mJy kpc$^2$ for Magellanic Cloud pulsars,
the effect of selection is less and the distribution approximates the
intrinsic luminosity function of pulsars, which in this plot of $d\log
N/d\log L$ has a slope of about $-1$ \citep{lor04}. Although we are
still limited by small-number statistics, there is no evidence that
the Magellanic Cloud pulsars have a different luminosity function from
their Galactic counterparts.

\section{Conclusions}
An extensive survey of the Large and Small Magellanic Clouds for radio
pulsars using the Parkes radio telescope at 1400 MHz has resulted in
the discovery of 14 pulsars, 12 of which are believed to be associated
with the Clouds. The average limiting flux density of the survey was
about 0.12 mJy. In general, the associated pulsars appear to be
located in the more central regions of each Cloud where there is
significant HI emission. Two long-period pulsars were discovered, one
of which appears to be a very old pulsar located in the Galactic halo;
the other is notable for its very strong implied surface dipole
magnetic field.

This survey brings the number of spin-powered pulsars believed to be
associated with the Magellanic Clouds to 20. Mean free electron
densities averaged over the volume of the Clouds are similar to those
in the disk of our Galaxy.  Observed luminosities are at the high end
of the luminosity function observed for Galactic pulsars but
consistent with it. Despite the observed pulse periods and ages
covering relatively wide ranges, there is little or no correlation of
radio luminosity with either of these parameters.

It will be difficult to significantly improve on this survey with
existing instrumentation as the required observation times are too
long to be feasible. Although searches for extra-galactic pulsars at
relatively low frequencies with large radio telescopes such as Arecibo
may have some success, any major increase in the their number will
most probably have to wait until the advent of instruments of much
larger collecting area, for example, the proposed Square Kilometer
Array.

\section*{Acknowledgments}
We thank our colleagues for assistance, especially with the timing
observations. The Parkes telescope is part of the Australia Telescope
which is funded by the Commonwealth Government for operation as a
National Facility managed by CSIRO.

%\bibliographystyle{apj} 
%\bibliography{journals,modrefs,psrrefs,crossrefs}

\end{document}